\documentclass[pra,aps,superscriptaddress,balancelastpage,twocolumn]{revtex4}
\usepackage{amsmath}
\usepackage{graphicx}
\usepackage{amsfonts}
\usepackage{amssymb}

\providecommand{\U}[1]{\protect\rule{.1in}{.1in}}
\providecommand{\U}[1]{\protect\rule{.1in}{.1in}}

\begin{document}

\title{Phase-Sensitive Enhanced Absorption, Transmission and Slow Light in a Cross-cavity Magnomechanical System}
\author{Amjad Sohail}
\email{amjadsohail@gcuf.edu.pk}
\affiliation{Department of Physics, Government College University, Allama Iqbal Road,
Faisalabad 38000, Pakistan}
\affiliation{Instituto de Física Gleb Wataghin, Universidade Estadual de Campinas, Campinas, SP, Brazil}
\author{Hazrat Ali}
\affiliation{Department Physics, Abbottabad University of Science and Technology, Havellian, 22500, KPK, Pakistan}
\author{K. B. Emale}
\affiliation{Department of Physics, Faculty of Science, University of Yaounde I, P.O.Box 812, Yaounde, Cameroon}
\author{Mohamed Amazioug}
\affiliation{LPTHE-Department of Physics, Faculty of Sciences, Ibnou Zohr University, Agadir, Morocco}
\author{Rizwan Ahmed}
\email{rizwanphys@gmail.com}
\affiliation{Physics Division, Pakistan Institute of Nuclear Science and Technology
(PINSTECH), P. O. Nilore, Islamabad 45650, Pakistan}

\begin{abstract}
We theoretically propose a scheme to explore the magnetically and magnomechanically induced transparency phenomena in a cross-cavity magnomechanical system, focusing on the role of relative phase and the intensity of the two probing fields in enhancing the absorption and transmission spectra and manipulating the group delay of the transmitted light. Interestingly, the relative phase of the two probe fields could have overwhelming effects on both the absorption spectrum and the group delay of the output field.
Tuning the relative phase and amplitude of the probe fields can suppress or enhance the absorption and transmission spectra. The combined effect of the magnon-photon and magnon-phonon couplings, along with relative phase modulations, helps to switch the probe field's behavior from subluminal to superluminal in the current system. 
The current study offers a straightforward and practical approach, demonstrating the capability to employ the relative phase for the modulation of microwave signals within the cavity magnomechanical system, providing insights for the design of information transduction and quantum sensing.
\end{abstract}

\maketitle

\section{Introduction}
The study of quantum effects at the macroscopic level has attracted much interest since the early days of quantum mechanics. In the present era, there are many platforms to study macroscopic quantum effects in macroscopic systems \cite{1}. These include quantum dots \cite{dot}, cold atoms \cite{cold}, trapped ions \cite{ion}, cavity optomechanics \cite{opt}, superconducting devices (e.g., SQUIDS) \cite{SC} and magnomechanical systems \cite{mmec}. Out of these many systems, cavity optomechanics and magnomechanics have gained considerable attention in recent years. These systems offer various effects that have been proposed theoretically and because of technological advancements, experimentally as well \cite{exp1}. Applications of these systems include the generation of macroscopic entanglement \cite{ent,ent1,ent2}, tests for the foundations of quantum mechanics \cite{found,found2}, optomechanical cooling \cite{cool}, quantum dense coding \cite{dcod,dcod1}, quantum sensing and metrology \cite{sens}, optomechanical (and magnomechanical) induced transparency, and many more \cite{trans}. 

Cavity magnomechanical systems have gained a lot of interest and emerged as an effective platform for macroscopic quantum effects. These systems are mainly based on microwave (MW) field(s) of a cavity coupled to magnons in a single-crystal yttrium iron garnet ($Y_{3}Fe_{5}O_{12}$; YIG) \cite{magn,magn1,magn2}. In the magnomechanical system, a magnetostrictive force acts like radiation pressure in a conventional optomechanical system and the YIG sphere can be affected by the applied magnetic field. There are various studies for the light-matter interaction that can be efficiently studied using magnomechanical systems \cite{LM,LM1,LM2}. This remarkable behavior is because of the unique properties of YIG since it has a high Curie temperature and spin density as well as a low decay rate \cite{curie,curie1}. Generally, the integration of magnonic elements with a conventional optomechanical system forms a magnomechanical system which, in recent years, provided an exceptional framework for the investigation of quantum features at the macroscopic scale.

Out of many interesting applications, Optomechanically induced transparency (OMIT) and absorption have been extensively studied in the recent past \cite{omit,omit1,omit2}. The efficiency of OMIT process is mainly based on the optical transmission and delay. In an interesting study \cite{omit}, Kronwald and Marquardt discussed the OMIT in the nonlinear regime where optomechanical interactions become significant. In addition to their theoretical proposal, they also reported experimental demonstration that pulsed transistor-like switching of transmission works even in this regime. More recently, much attention has been paid to photon-magnon hybridization, resulting in a magnetically induced transparency (MIT) phenomenon at room temperature \cite{m1,m2,m3,m4}. The special characteristics of the YIG sphere make it a perfect mechanical oscillator, generating magnon–phonon coupling and, hence uncovering the vital phenomena of magnomechanically induced transparency (MMIT) \cite{magn}. In such a scheme, MMIT emerges from magnon-phonon-photon interaction via dipole and magnetostrictive interactions, respectively. In addition to MMIT, there is another interesting feature of slow/fast light in the magnomechanical system which is due to the rapid change of the group index in the neighborhood of resonance \cite{slow,slow2,slow3}.  In their seminal paper, Li \textit{et. al.}, 
proposed a scheme to control the dynamics of MMIT and group delay by choosing the relative phase of applied fields in a magneto-mechanical system \cite{phase}. In their scheme, the phase can be adjusted by taking a minimal value of the magnetic field of the magnon drive. However, we show that the cavity drive field can also control the phase.
We explore phase-sensitive double MIT and double MMIT in the feasible working experimental regime. In addition, as a natural consequence of the tunable double windows of MMIT, we also present results for slow and fast light via group delay. Interestingly, this phenomenon is not only phase sensitive but also gives a larger range of switchable group delay, which shows that our system is more controllable. In this way, we can assert that the MMIT in our case might find many applications in optical switching, quantum devices, and quantum information processing protocols \cite{app}.

The organization of the paper is as follows. In Sect. 2, we present the model and the Hamiltonian of the cross-cavity magnomechanical system. We also describe the dynamical equations of the magnomechanical system using the quantum Langevin approach in this section. Section 3 discusses the results obtained for the double MIT and MMIT. The enhanced transmission spectra of the probe field and the tunable group delays are shown in Section 4. Finally, we conclude the study in Sect. 5.

\section{The Model}
As illustrated in Fig. 1, the magnomechanical system under discussion
consists of a YIG sphere and two cross-shaped microwave cavities. YIG sphere is subjected to a uniform bias magnetic field ($z$-direction), which excites the magnon modes inside it. These modes are then connected to both cavity modes through magnetic-dipole interactions. Furthermore, the lattice structure of the YIG spheres is deformed as a result of a
fluctuating magnetization caused by the excitation of the magnon modes. As a result, the magnetostrictive interaction which is responsible for the deformation of the geometry of the YIG sphere, establishes the interactions
between magnon and phonon. The single-magnon magnomechanical coupling strength, which is assumed to be very weak, depends on the diameter of the YIG sphere and the direction of the
external bias field. However, we consider a strong
external microwave drive that drives the magnon mode of the YIG sphere. In
our model, this microwave drive acts as a control field and strengthens the magnon-phonon interaction inside the YIG sphere. Furthermore, each cavity is driven by a weak probe field, however, there exists a relative phase $\phi$ between the two fields, with amplitude $\varepsilon _{x,y}=\sqrt{2k_{x,y}\wp_{x,y}/\hbar\omega_{x,y}}$. In the current study, we consider a
high-quality YIG sphere composed of ferric ions Fe$^{+3}$ of density $\rho
=4.22\times 10^{27}$m$^{-3}$ and diameter $D=250\mu $m. This results in a
total spin $S=\frac{5}{2}\rho V=7.07\times 10^{14}$m$^{-3}$, where $V$ is
the volume of the YIG sphere.
The Hamiltonian of the system is given by
\begin{eqnarray}
\hat{H}/\hbar  &=&\sum\limits_{k=1}^{2}\omega _{k}c_{k}^{\dagger
}c_{k}+\omega _{m}m^{\dagger }m+\omega _{b}b^{\dagger }b  \notag \\
&&+\sum\limits_{k=1,2}\Gamma _{k}\left( c_{k}m^{\dagger }+c_{k}^{\dagger
}m\right) +g_{mb}m^{\dagger }m\left( b^{\dagger }+b\right)   \notag \\
&&+i\left( \varepsilon _{m}m^{\dagger }e^{-i\omega _{d}t}-\varepsilon
_{m}^{\ast }me^{i\omega _{d}t}\right)   \notag \\
&&+i\left( c_{1}^{\dagger }\varepsilon _{x}e^{-i\omega
_{x}t}+c_{1}\varepsilon _{x}^{\ast }e^{i\omega _{x}t}\right)   \notag \\
&&+i\left( c_{2}^{\dagger }\varepsilon _{y}e^{-i\omega _{y}t}e^{i\phi
}+c_{2}\varepsilon _{y}^{\ast }e^{i\omega _{y}t}e^{-i\phi }\right).
\end{eqnarray}%
The first three terms in Eq. (1) reflect the free Hamiltonian of the cavity
modes, magnon mode, and phonon mode, respectively. Here, $c_{k}^{\dagger }(c_{k})$, $%
m^{\dagger }(m)$, and $b^{\dagger }(b)$ are the creation (annihilation)
operators of the respective cavity mode, the magnon mode, and the phonon
mode, respectively. Furthermore, $\omega _{k}$, $\omega _{m}$, and $\omega
_{b}$, represent the respective resonance frequencies of the cavity, magnon,
and phonon modes. It is worth mentioning that the operators $m^{\dag }$ and
$m$ are the bosonic field operators for the magnon mode and its frequency can be
determined employing the gyromagnetic ratio $\gamma _{g}$ and the bias
magnetic field, $H$, via $\omega _{m}=\gamma _{g}H$. The fourth term
represents the interaction between the magnon modes and the $k$th cavity with the 
optomagnonical coupling strength $\Gamma _{k}$. The fifth term denotes the
interaction between the magnon and modes with the magnomechanical coupling $%
g_{mb}$. The last three terms are input-driving field terms. The Rabi
frequency $\varepsilon _{m}=\frac{\sqrt{5N}}{4}\gamma _{g}H_{d}$ indicates
the strength of the coupling between the driving field of the microwave and
the magnon, where $N=\rho V$ stands for the total spin number of the YIG crystal.
\begin{figure}[tbp]
\centering
\includegraphics[width=1\columnwidth,height=1in]{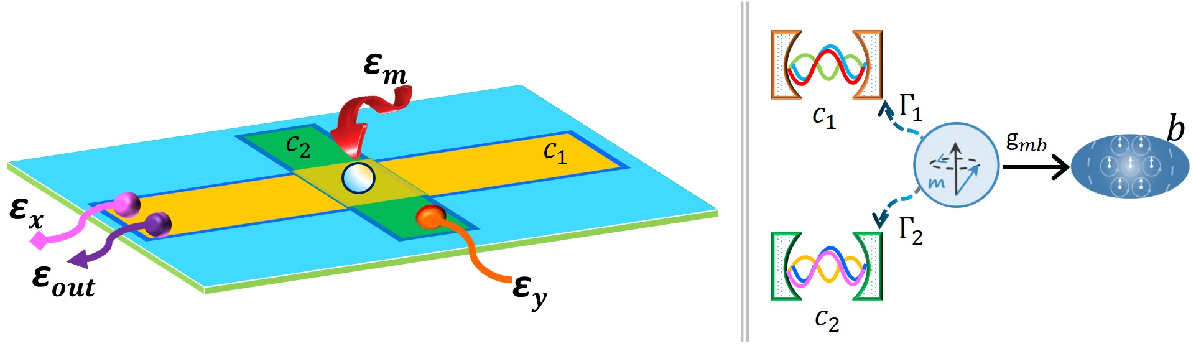} \centering
\caption{Schematic diagram of a cross-cavity magnomechanical system in which a magnon mode is simultaneously coupled to phonon mode as well as to both the cavity modes.}
\end{figure}
The total Hamiltonian for the current system under the rotating wave
approximation at the driving field frequency $\omega _{d}$ is given by
\begin{eqnarray}
\hat{H}/\hbar  &=&\sum\limits_{k=1}^{2}\Delta _{k}c_{k}^{\dagger
}c_{k}+\Delta _{m}^{0}m^{\dagger }m+\omega _{b}b^{\dagger }b  \notag \\
&&+\sum\limits_{k=1,2}\Gamma _{k}\left( c_{k}m^{\dagger }+c_{k}^{\dagger
}m\right) +g_{mb}m^{\dagger }m\left( b^{\dagger }+b\right)   \notag \\
&&+i\varepsilon _{m}\left( m^{\dagger }-m\right) +i\left( c_{1}^{\dagger
}\varepsilon _{x}e^{-i\delta _{x}t}+c_{1}\varepsilon _{x}^{\ast }e^{i\delta
_{x}t}\right)   \notag \\
&&+i\left( c_{2}^{\dagger }\varepsilon _{y}e^{-i\delta _{y}t}.e^{i\phi
}+c_{2}\varepsilon _{y}^{\ast }e^{i\delta _{y}t}e^{-i\phi }\right).
\end{eqnarray}%
Here $\Delta _{k}=\omega _{k}-\omega _{d}$ ($k=1,2$), $\Delta
_{m}^{0}=\omega _{m}-\omega _{d}$, and $\delta _{j}=\omega _{j}-\omega _{d}$
($j=x,y$) represent the frequency detunings
of the cavity mode, magnon mode, and probe from the driving field. To understand the dynamics of the system within the semiclassical
limit, we can write the Heisenberg-Langevin equations
\begin{eqnarray}
\dot{c_{1}} &=&-\left( i\Delta _{x}+\kappa _{x}\right) c_{1}-i\Gamma
_{1}m+\varepsilon _{x}e^{-i\delta _{x}t},  \notag \\
\dot{c_{2}} &=&-\left( i\Delta _{y}+\kappa _{y}\right) c_{2}-i\Gamma
_{2}m+\varepsilon _{y}e^{-i\delta _{y}t}e^{i\phi },  \notag \\
\dot{m} &=&-\left( i\Delta _{m}^{0}+\kappa _{m}\right)m \notag \\
&&-i\sum\limits_{j=1,2}\Gamma _{k}c_{k}-ig_{mb}m\left( b^{\dagger }+b\right)
+\varepsilon _{m},  \notag \\
\dot{b} &=&-i\omega _{b}b-ig_{mb}m^{\dagger }m-\gamma _{b}b.  \label{LAEQ}
\end{eqnarray}%
Here, we have omitted the thermal and quantum input noise terms because we are
interested in investigating the mean response of the current system to the
applied probing field. Within the semiclassical perturbation framework, we
assume that the probe microwave field is substantially weaker than the
control microwave field. Consequently, we can expand each operator as $%
z=z_{s}+\delta z$, where $z_{s}$ ($\delta z$) is the steady-state value
(small fluctuation) of the operator. First, we consider the steady-state
solutions, which are given by
\begin{eqnarray}
b_{s} &=&\frac{-ig_{mb}}{i\omega _{b}+\gamma _{b}}\left\vert
m_{s}\right\vert ^{2},  \notag \\
c_{j,s} &=&\frac{-i\Gamma _{j}m_{s}}{\kappa {j}+i\Delta _{j}},  \notag
\\
m_{s} &=&\frac{\Omega \zeta _{x}\zeta _{y}}{\zeta _{x}\zeta _{y}\zeta
_{m}+\Gamma _{1}^{2}\zeta _{y}+\Gamma _{2}^{2}\zeta _{x}},  \label{MAV}
\end{eqnarray}%
where $\zeta _{s}=\kappa _{s}+i\Delta _{s}$ ($s=x,y,m$) and $\Delta
_{m}=\Delta _{m}^{0}+g_{mb}(b_{s}+b_{s}^{\ast })$. We assume that the current system is in a resolved sideband regime, in which $\omega_{b}\gg\kappa_m,\kappa_j$. In this regime, we can safely take $\Delta_{x,y}=\Delta_{m}=\omega_{b}$. Furthermore, Eq. (\ref{LAEQ}), can be easily solved by introducing the slowly varying operators as $\delta c_{k}=\delta c_{k}e^{-i\Delta _{k}t}$, $\delta m=\delta me^{-i\omega _{\Delta _{m}}t}$, and $\delta b=\delta be^{-i\omega _{b}t}$.
The amplitude of the probe field is assumed to be significantly weaker than the coupling of the external microwave drive on magnon mode. By taking
into account, the perturbation caused by the input probe field up to the
first-order term, we obtain the set of linearized equations of motion
\begin{eqnarray}
\delta \dot{c_{1}} &=&-\kappa _{x}\delta c_{1}-\iota \Gamma _{1}\delta
m+\varepsilon _{x}e^{-i\delta },  \notag \\
\delta \dot{c_{2}} &=&-\kappa _{y}\delta c_{2}-\iota \Gamma _{2}\delta
m+\varepsilon _{y}e^{-i\delta }e^{\iota \phi },  \notag \\
\delta \dot{m} &=&-\kappa _{m}-\iota \Gamma _{1}\delta c_{1}-\iota \Gamma
_{2}\delta c_{2}-\iota G_{mb}\delta b, \notag \\
\delta \dot{b} &=&-\gamma _{b}\delta b-\iota G_{mb}^{\ast }\delta m,
\end{eqnarray}%
where $G_{mb}= g_{mb}m_s$ is the effective magnomechanical coupling coefficient. Note that for a fixed $g_{mb}$, the value of $G_{mb}$ can be modified/enhanced via $m_s$ by an external magnetic field (see Eq. (\ref{MAV})). In addition, we have assumed 
$\varepsilon _{x}=\varepsilon _{y}=\varepsilon _{p}$, 
$\delta _{x}=\delta _{y}=\delta $, and $\sigma =\delta
-\omega _{b}$ is the effective detuning. To solve the above set of linearized
equations, we apply an ansatz $\delta z=z_{+}e^{-i\sigma}+z_{-}e^{i\sigma}$
where the coefficients $z_{+}$ and $z_{-}$ (with $z=c_{1},c_{2},m,b$), respectively, correspond to the components at the frequencies $\omega_p$ and $2\omega_d-\omega_p$.
Then it is straightforward to obtain the final solution at the probe frequency
\begin{eqnarray}
c_{1+}&=&\frac{\Lambda}{\alpha _{1}\Lambda +\Gamma _{1}^{2}\alpha _{2}\alpha _{b}} -\frac{\left[\Gamma _{1}\Gamma
_{2}\alpha _{b}\right]\xi e^{\iota \phi }}{\alpha _{1}\Lambda +\Gamma _{1}^{2}\alpha _{2}\alpha _{b}}, \label{OUT}
\end{eqnarray}%
where $ \Lambda=\alpha _{2}\alpha _{m}\alpha _{b}+\Gamma _{2}^{2}\alpha
_{b}+\left\vert G_{mb}\right\vert ^{2}\alpha _{2}$, $\xi =\frac{\varepsilon _{x}}{\varepsilon _{y}}$ and $\alpha _{z}=\kappa
_{z}-i\sigma $ ($z=x,y,m,b$). In Eq. (\ref{OUT}), the first term is the contribution of the double MMIT, and the second term incorporates the effect resulting from the phase difference between two weak driving fields.
Based on the input-output theory $\varepsilon_{T}=\varepsilon_{in}-\kappa_{1}c_{1+}$, we can write the equation for the amplitude of the output field at the probe frequency, is given by
\begin{equation}
\varepsilon_{T}=\frac{2\kappa_1 c_{1+}}{\varepsilon _{p}}=\chi_{r}+i\chi_{i}.
\end{equation}
It is vital to mention that $\varepsilon_{T}$ is a complex quantity. In addition,
the real and imaginary parts of $\varepsilon_{T}$ exhibit the absorption (in-phase) and dispersion (out-of-phase) spectrum of the output field quadratures at the probe frequency.
\section{Magnetically and Magnomechanically induced transparency window Profiles \label{secM}}
This section explicitly shows the viability of MIT and MMIT in our system.
We utilize the following parameters from a recent experiment on a hybrid magnomechanical system for numerical computation. $%
\omega_{1}=\omega_{2}=\omega=2\pi\times 10$ GHz, $\omega_{b}=2\pi \times 15$ MHz, $\kappa_{x}=2\pi \times 2.1$ MHz, $\kappa_{y}= 2\pi \times 0.15$ MHz, $\kappa_{m}=2\pi \times0.1$ MHz,
$\Gamma_{1}=\Gamma_{2}=\Gamma=2\pi \times 3.2$ MHz, $\gamma_{b}=10^{-5}\omega_{b}$, $g_{mb}=2\pi \times 0.3$ Hz, $T=10$ mK, $H_d=1.3\times10^{-4}, \gamma_{G}/2\pi=28$ GHz/T, $D=125\mu$ m, and $\rho=4.22\times10^{27}$ \cite{magn,phase}.
Firstly, we demonstrate the physics behind the double transparency by ignoring the phase effect but considering the key role of different couplings present in our system. Figure 2 (a) shows the absorption spectrum of the output field at the probe frequency for different coupling strengths. It can be seen that if we ignore both the magnetostrictive interaction and magnetic dipole interaction (with cavity 2), we obtained a magnetically induced transparency (MIT) profile (as shown by the solid blue line), where a standard Lorentzian curve splits at the opacity point and bilateral symmetric absorption peaks appear at either sideband, i,e., $\sigma=\pm 0.32\omega_b$. Obviously, by tuning the magnon-photon coupling (with cavity 1) and the microwave driving field power, the width of such a transparency window can be adjusted. We observe the double MMIT windows in case of taking either magnetostrictive interaction or magnetic dipole interaction (with cavity 2) as shown by the dot-dashed green line and solid yellow line. However, one can note that the role of the two couplings is slightly different. If we take $\Gamma_{1}$ and $G_{mb}$ ($\Gamma_{2}=0$), the height of the middle peak (two side peaks) of the double-window profile increases (unchanged) and the width of the two windows widened. On the other hand, taking $\Gamma_{1}$ and $\Gamma_{2}$ ($G_{mb}$=0), not only the width of the side peaks is increased with slight sharpness of the middle peak at the opacity point. We observe combined effects if we include all the couplings present in the system. In principle, we must obtain three roots, but due to a very small value of $\gamma_b$, its effect can only be observed at the opacity point, which can be seen by the inset in Fig. 2(a). Hence, we can see exactly the location of the peaks of the two MIT windows at $\sigma=\pm (G_{mb}+\Gamma)=\pm(0.32+0.18)=\pm 0.49\omega_b$.
\begin{figure}[tbp]
\centering
\includegraphics[width=1\columnwidth,height=3.5in]{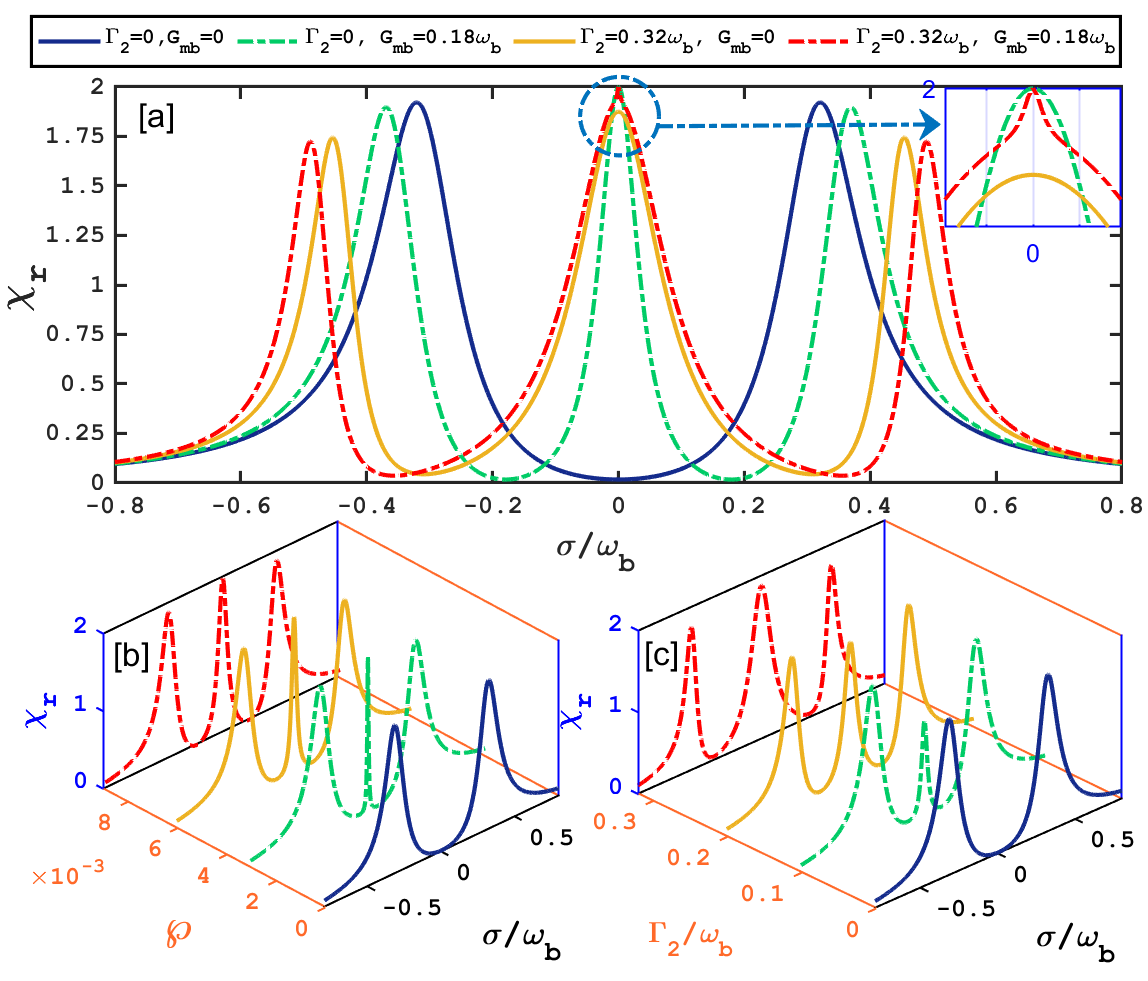} \centering
\caption{(a) The response of MMS as a function of the probe detuning.$(\protect%
\xi=0)$. (b) The absorption spectrum of the (a) MIT ($G_{mb}=0$, $\Gamma_{2}\neq0
$)and (c) MMIT ($G_{mb}\neq0$, $\Gamma_{2}=0$) windows profile against the
normalized probe field detuning. $(\protect\xi=0)$.}
\end{figure}
\begin{figure}[tbp]
\centering
\includegraphics[width=1\columnwidth,height=4in]{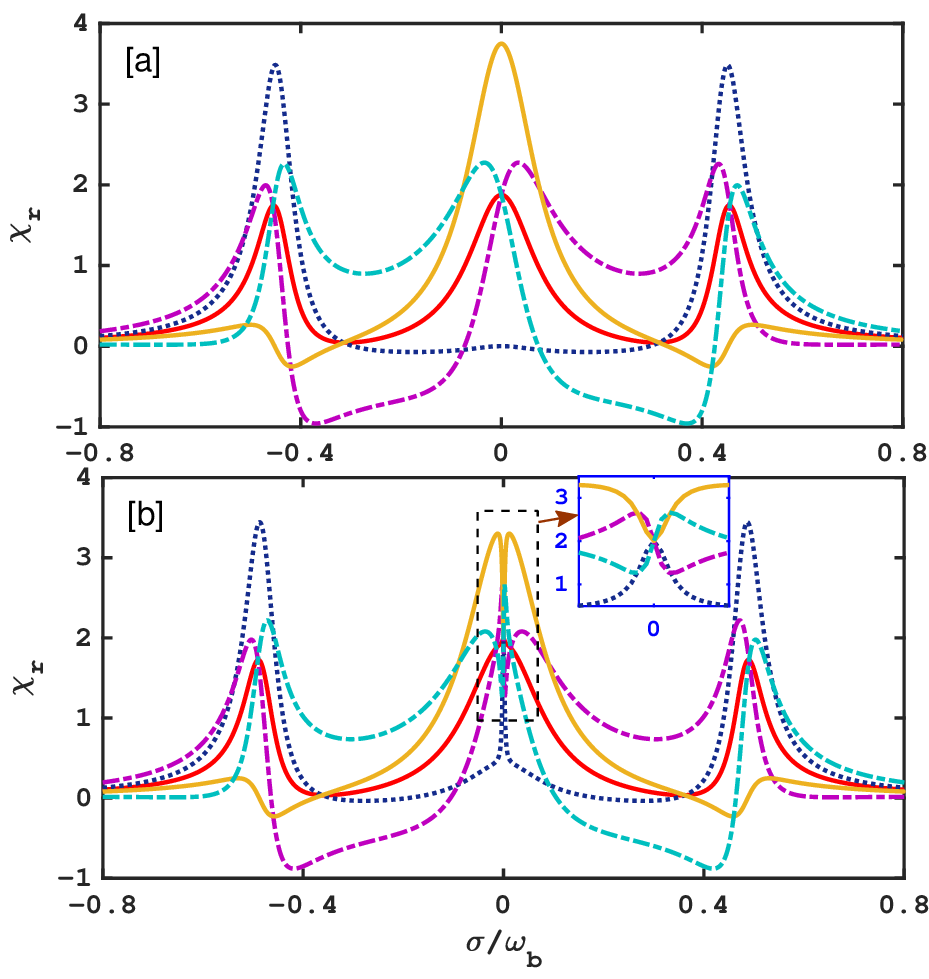} \centering
\caption{The absorption spectrum of the (a) MIT ($G_{mb}=0$, $\Gamma_{2}\neq0
$)and (b) MMIT ($G_{mb}\neq0$, $\Gamma_{2}\neq0$) windows profile against
the normalized probe field detuning for different values of phase angle $%
\protect\xi=0$ (blue curve). For the rest of the curves $\protect\xi=1$, $%
\protect\phi=0$ (blue dotted  curve), $\protect\phi=\protect\pi/2$(purple dot-dashed curve), $%
\protect\phi=\protect\pi$ (orange solid curve), and $\protect\phi=3\protect\pi/2$ 
(cyan dot-dashed curve).}
\end{figure}

As we turn on the magnon-phonon coupling ($G_{mb}$) while keeping $\Gamma_{2}=0$, we see two transparency windows in the absorption spectrum. The single MIT window splits into the double window as shown in Fig. 2(b) because of the nonzero magnetostrictive interaction. Hence, we can safely say that the right transparency window appears because of the magnon-phonon interaction, which we call the MMIT window. Two dips in the absorption spectrum in Fig. 2(b) correspond to two transparency frequencies of the probing field, i.e., the energy levels are split into two levels by $\pm G_{mb}$ and hence configure two MIT energy-level configurations, that is, the dips of the two MIT windows can be exactly located at $\sigma=\pm G_{mb}$. In another case, we can achieve the double window profile by switching off the magnon-phonon coupling ($G_{mb}$) and taking the nonzero couplings between the magnon mode and cavity 2 as shown in Fig. 2(c). In this situation, the energy levels are divided into two sub-levels, and the dips of the two MIT windows can be exactly located at $\sigma=\pm \Gamma$. Thus, our system has two ways to generate a double transparency window profile.
\begin{figure}[tbp]
\centering
\includegraphics[width=1\columnwidth,height=2.7in]{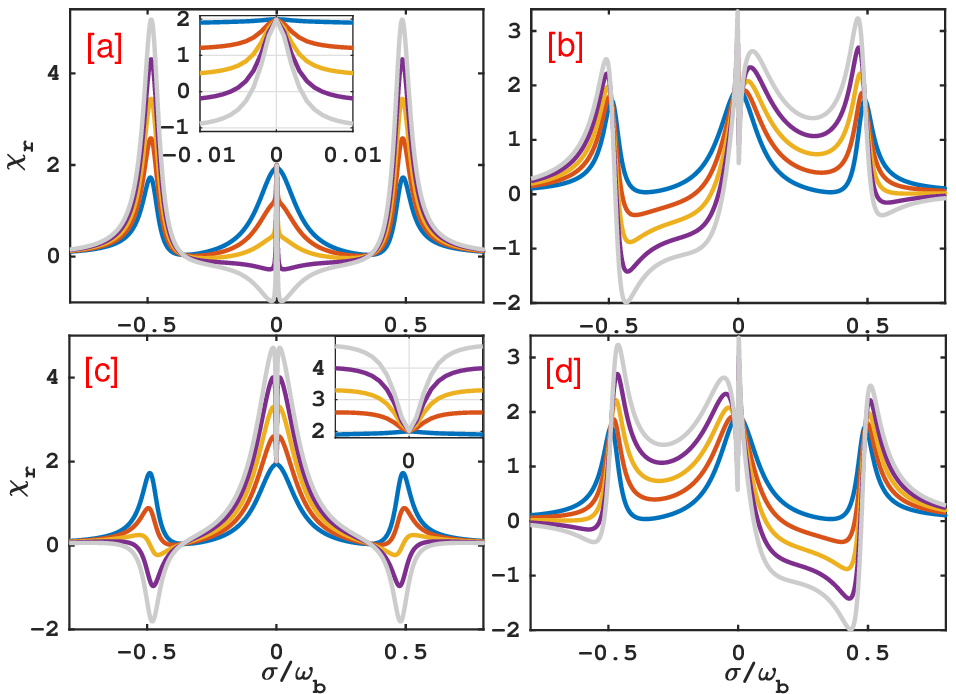} \centering
\caption{The absorption spectrum of the MMIT windows profile against the
normalized probe field detuning for different value of $\protect\xi$ when
(a) $\protect\phi=0$ (b) $\protect\phi=\protect\pi/2$ (c)$\protect\phi=%
\protect\pi$, and (d) $\protect\phi=3\protect\pi/2$. Here $\protect\xi=0$
(blue curve), $\protect\xi=0.5$ (orange curve), $\protect\xi=1$ (yellow
curve), $\protect\xi=1.5$(violet curve), and $\protect\xi=2$(light gray
curve).($G_{mb}\neq0$, $\Gamma_{2}\neq0$).}
\end{figure}
\begin{figure}[b!]
\centering
\includegraphics[width=1\columnwidth,height=1.6in]{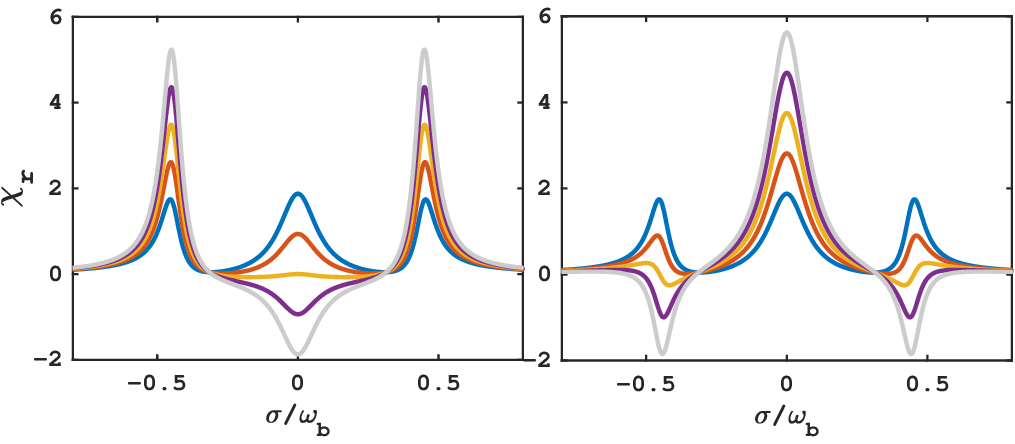} \centering
\caption{The absorption spectrum of the MMIT windows profile against the
normalized probe field detuning for different value of $\protect\xi$ when
(a) $\protect\phi=0$, and (b)$\protect\phi=\protect\pi$. Here $\protect\xi=0$
(blue curve), $\protect\xi=0.5$ (orange curve), $\protect\xi=1$ (yellow
curve), $\protect\xi=1.5$(violet curve), and $\protect\xi=2$(light gray
curve).($G_{mb}=0$, $\Gamma_{2}\neq0$).}
\end{figure}

In the rest of the paper, our main focus is analyzing the phase effect and the amplitude ratio between the two probe fields on the output field's response. In Fig. 3 (a), we plot the absorption spectrum $\chi_r$ as a function of $\sigma/\omega_b$ in the absence of magnomechanical interaction, i.e., $G_{mb}=0$, and by tuning the phase, keeping the amplitude ratio of the two probing fields fixed.
For simplicity, we have considered the ratio of amplitude between the two probe fields $\xi=1$. When $\phi= 0$, the interference between the two terms in Eq. (\ref{OUT}) yields an enhanced absorption of the two side peaks, while the complete suppression of absorption at the point $\sigma=0$. In contrast, the absorption of the central peak at the point $\sigma=0$ enhanced, letting the side peaks into amplification for $\phi= \pi$. Furthermore, when $\phi=\pi/2$ ($\phi=3\pi/2$), we obtained the asymmetric profile representing the enhanced (diminished) absorption of the right (left) window. Therefore, by tuning the phase, the absorptive
behavior can be modulated from the sideband regions, and absorption and amplification can be obtained simultaneously. In Fig. 3(b), we repeat the same process, taking both $G_{mb}$ and $\Gamma_{2}$. We obtained enhanced absorption of the two side peaks, while the middle peak became shallower for $\phi= 0$. Note that an enhanced MMIT appears around the opacity point for $\phi= \pi$. Hence, by tuning the phase, we can switch the output probe's response between the transparency and absorption states at $\sigma=0$. Furthermore, the behavior for $\phi=\pi/2$ and $\phi=3\pi/2$ is almost the same as in the previous case, except for the dispersive-type response around the opacity point.

To further explore the system's response more clearly, we plot the absorption spectrum as a function of probe detuning for different values of $\xi$ with the fixed value of the phase angles. To see the effect of amplitude ratio $\xi$ more clearly, we plot the absorption spectra versus normalized probe
detuning with different $\xi$ in Fig. 4.
When $\phi=0$ ($\phi=\pi$),
the bilateral symmetric absorption peaks increase (decrease) monotonically and simultaneously decrease (increase) around the opacity point by increasing the value of $\xi$ as shown in Fig. 4(a) (Fig. 4(c)).
Furthermore, it can be seen from Fig. 4(b) (Fig. 4(d)) that the absorption of the central peak and the right (left) valley of the double MMIT window enhances as the amplitude ratio of the two probes increases when $\phi=\pi/2$ ($\phi=3\pi/2$). In Fig. 5, we adopt the same procedure as for Fig. 4 but without considering the magnomechanicl interaction. By comparing the two figures, one can see that we obtained almost the same behavior except for the variations around the opacity point.
\begin{figure}[tbp]
\centering
\includegraphics[width=1\columnwidth,height=1.3in]{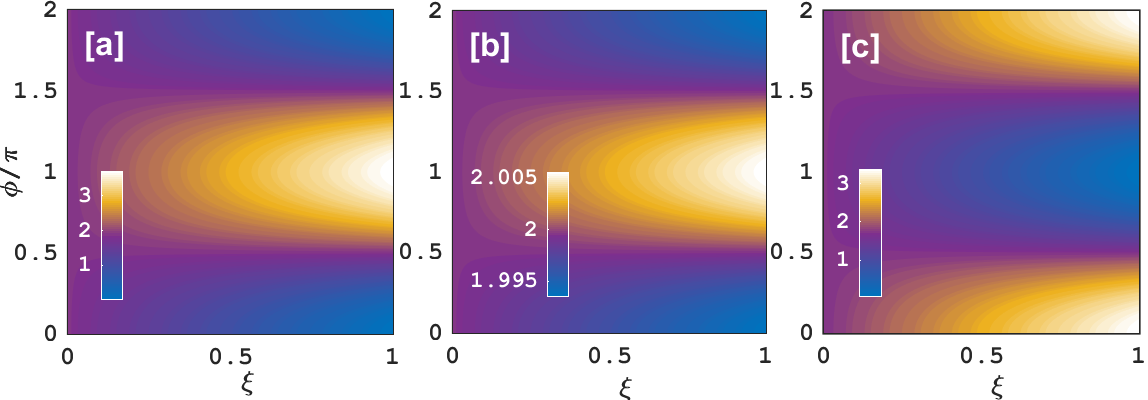} \centering
\caption{Contour plot of the absorption spectrum of the MMIT windows profile
against the phase angle $\protect\phi$ and the ratio of the intensities $%
\protect\xi$ when (a-b) $\sigma=0$ and (c) $\sigma=\pm0.49\omega$. Here (a) $G_{mb}=0$, $%
\Gamma_{2}\neq0$) and (b)(c) $G_{mb}=\neq$, $\Gamma_{2}\neq0$).}
\end{figure}

Next, we show how the amplitude ratio $\xi$ and the relative phase between two weak probe driving fields can effectively modulate absorption. The signature of the output field reflects the significance of the relative phase. In Fig. 6, we plot the absorption as a function of the amplitude ratio $\xi$ and the phase difference of the two weak probe driving fields at the opacity point and sideband regions. It is crucial to note that when the driving field's amplitude grows, the relative phase effect in the absorption spectra becomes more pronounced around the opacity point and the blue(red)-detuned area.
The absorption spectra at the opacity point, in the absence of magnomechanical interaction, are shown to reach the minimum (maximum) around $\phi=0$ ($\phi=\pi$) as shown in Fig. 6 (a). One can notice that the magnomechanical interaction weakens the phase effect at the opacity point, as shown in Fig. 6 (b). However, the weakening effect around the opacity point is much less than what we have already seen for the sideband regions.
Furthermore, we obtained a converse behavior around the sideband regions where the absorption spectra are shown to reach the maximum (minimum) around $\phi=0$ ($\phi=\pi$), see Fig. 6(c). Hence, magnomechanical interaction weakens the output field near the opacity point while enhancing around the sideband regions.
\section{Enhanced Transmission and Tunable SLOW/FAST LIGHT \label{secM}}
In this section, we study the transmission and group delay of the output response of the current cross-cavity magnomechanical system by investigating the relative phase of the probing fields on the transmission spectrum.
From Eq. (6), the transmission rate of the
system can be derived as
\begin{eqnarray}
T_p&=&1-\frac{\kappa_{1}c_{1+}}{\varepsilon_{p}e^{-i\phi}}, \notag \\
&=&T_m+T_{ph},
\end{eqnarray}
where
$T_m=1-\frac{\Lambda}{\alpha _{1}\Lambda +\Gamma _{1}^{2}\alpha _{2}\alpha _{b}}$, and $T_{ph}= -\frac{\left[\Gamma _{1}\Gamma
_{2}\alpha _{b}\right]\xi e^{\iota \phi }}{\alpha _{1}\Lambda +\Gamma _{1}^{2}\alpha _{2}\alpha _{b}}$.

\begin{figure}[tbp]
\centering
\includegraphics[width=1\columnwidth,height=4in]{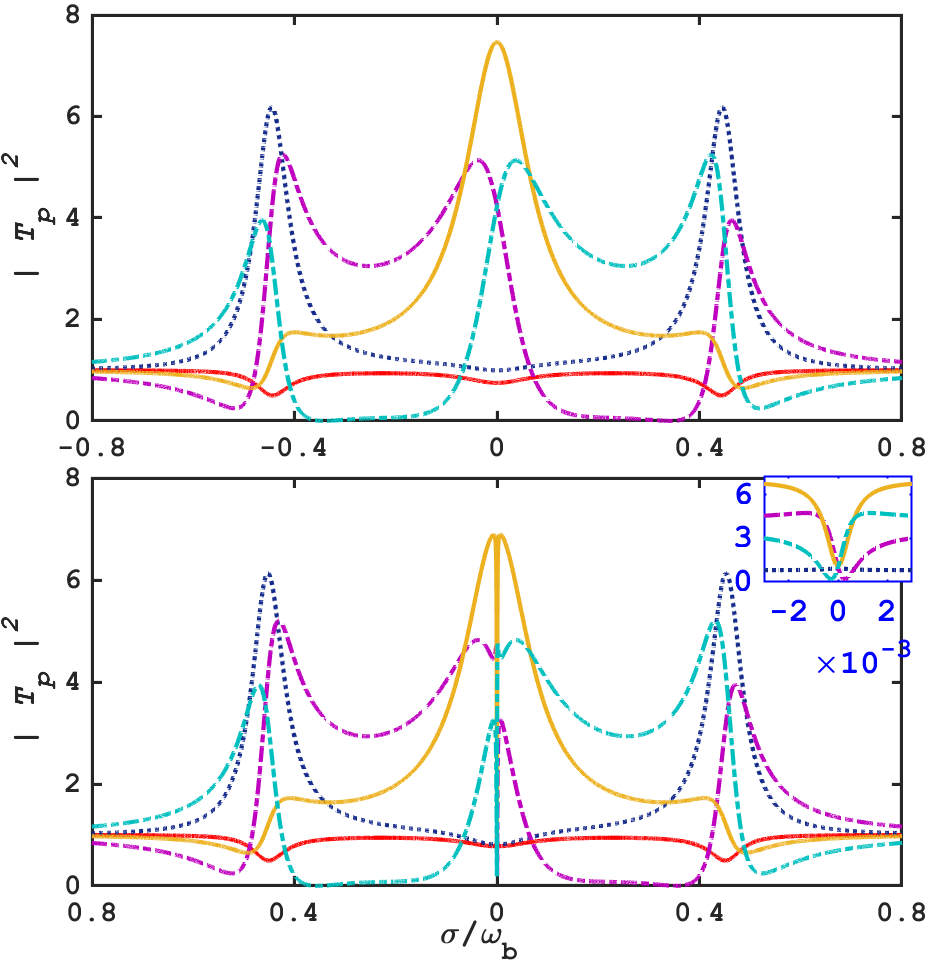} \centering
\caption{The transmission spectrum $|T_p|^2$ of the (a) MIT ($G_{mb}=0$, $\Gamma_{2}\neq0
$)and (b) MMIT ($G_{mb}\neq0$, $\Gamma_{2}\neq0$) windows profile against
the normalized probe field detuning for different values of phase angle $%
\protect\xi=0$ (blue curve). For the rest of the curves $\protect\xi=1$, $%
\protect\phi=0$ (blue dotted  curve), $\protect\phi=\protect\pi/2$(purple dot-dashed curve), $%
\protect\phi=\protect\pi$ (orange solid curve), and $\protect\phi=3\protect\pi/2$ 
(cyan dot-dashed curve).}
\end{figure}

In order to see the effect of the phase of the two probe fields on the probe transmission signal $|T_p|^2$, we plot the transmission spectrum against the normalized detuning for different phase angles in the absence of magnomechanical interaction. The two transmission valleys around the sideband regions (when $\xi=0$) transform into two enhanced peaks for $\xi\neq0$, $\phi=0$, hence we can say that the probe transmission of the output field varies from almost suppressed to greatly amplified signal around the sideband regions. Furthermore, probe transmission is greatly improved around the opacity point for $\phi=\pi$. Note that this enhancement extends the domain of the probe transmission over a wide range of normalized detuning values. Furthermore, it can be shown that there exists a significant enhancement in the transmission from the red (blue) detuned region for $\phi=\pi/2$ ($\phi=3\pi/2$). Fig. 7(b) shows almost similar behavior except for the variation at the opacity point. 
\begin{figure}[tbp]
\centering
\includegraphics[width=1\columnwidth,height=3in]{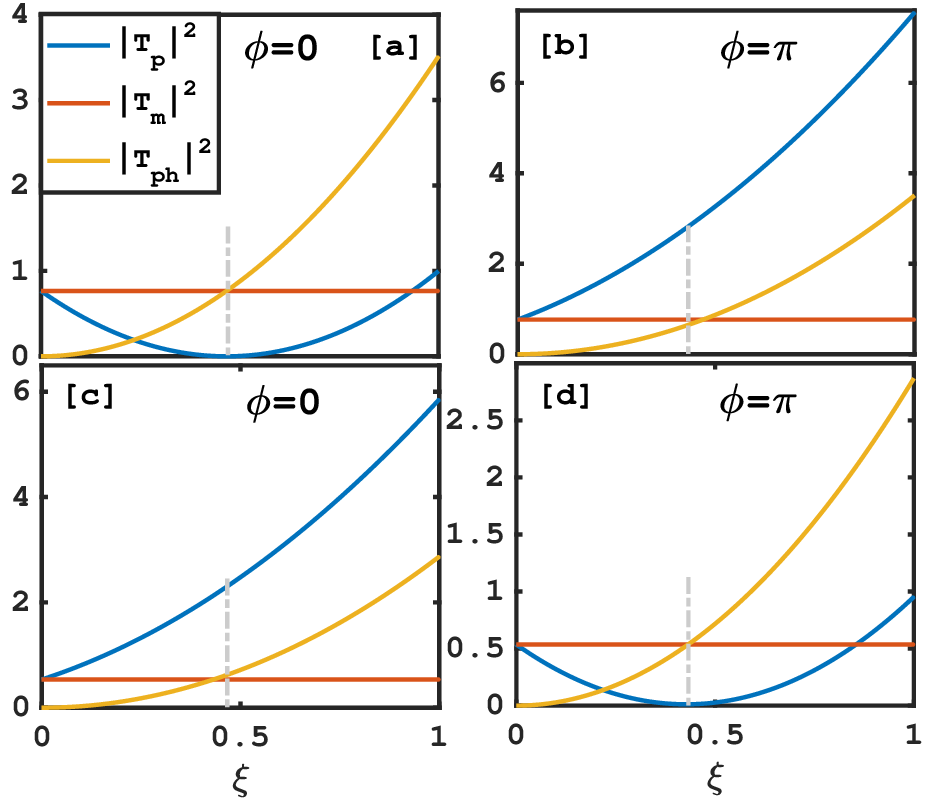} \centering
\caption{Plots of $|T_p|^2$, $|T_m|^2$, and $|T_{ph}|^2$ as a function of $\xi$ for (a-b) $\sigma=0$, $G_{mb}=0$, $\Gamma_{2}\neq0$, and (c-d) $\sigma=\pm0.49\omega_b$, $G_{mb}\neq0$, $\Gamma_{2}\neq0$. The other parameters are same in Fig. 7.}
\end{figure}

To shed additional light on the origin of the suppressing and enhancing of the output field induced by the phase effect, we plot $|T_P|^2$, $|T_m|^2$, and $|T_{ph}|^2$ as
functions of $\xi$ in Fig. 8. First we discuss the case where the magnomechanical interaction is ignored. In Fig. 8(a), it can be seen that there is destructive interference between $T_m$ and $T_{ph}$, and the strongest destructive interference can be seen when $T_P=0$ around $\xi=0.45$. When $\phi=\pi$, it can be seen that $T_{ph}$ rises monotonically with the enhancement of $\xi$. In contrast, $T_m$ remains constant. Therefore, we can safely say that there is constructive interference between $T_m$ and $T_{ph}$. We can observe a perfect constructive interference around $\xi=0.45$, which leads to $|T_P|^2\simeq4|T_m|^2\simeq4|T_{ph}|^2$.
Now we discuss the case with the magnomechanical interaction. It is crucial to mention that we do not observe any destructive interference between $T_m$ and $T_{ph}$ around the opacity point when we include the magnomechanical interaction. However, perfect constructive (destructive) interference is obtained for $\phi=0$ ($\phi=\pi$) around $\xi=0.45$. Complete constructive interference between $T_m$ and $T_{ph}$ results in the outcome $|T_P|^2\simeq4|T_m|^2\simeq4|T_{ph}|^2$ when $\phi=\pi$.

\begin{figure}[tbp]
\centering
\includegraphics[width=1\columnwidth,height=2.4in]{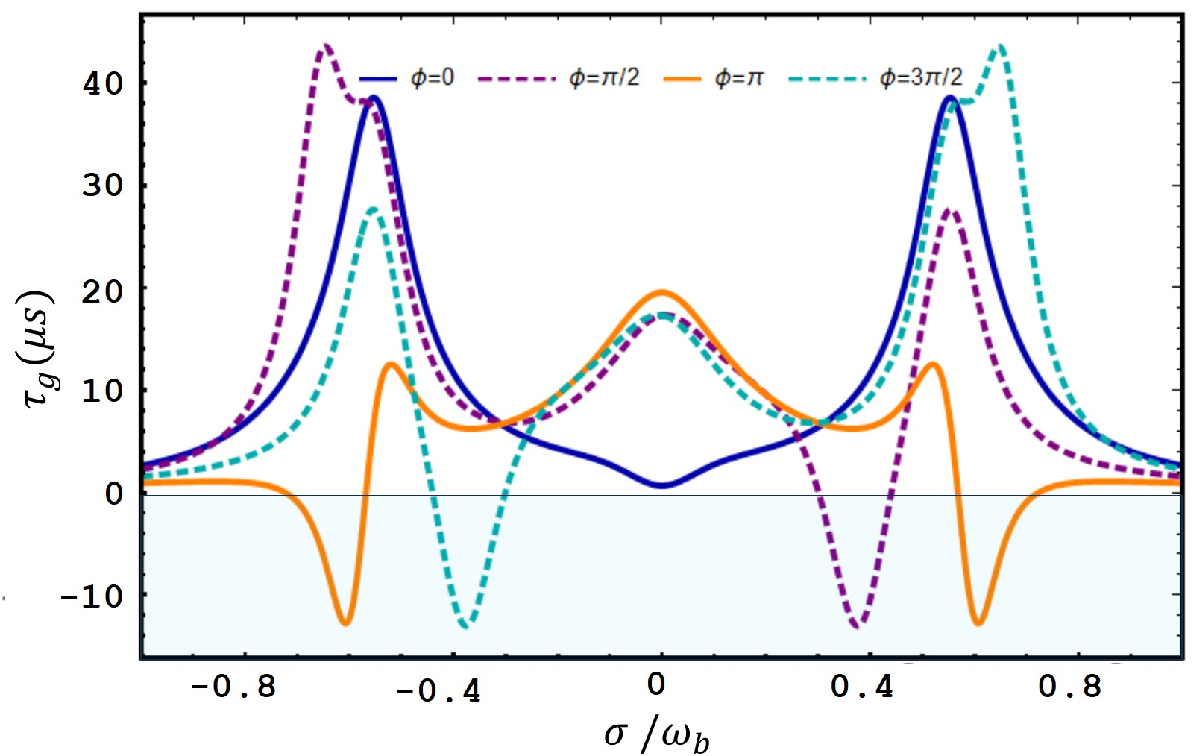} \centering
\caption{The group delay as a function of normalized probe detuning with the different phase angle $\protect\phi$. We take $\xi=1$, $\protect\phi=0$ (blue solid  curve), $\protect\phi=\protect\pi/2$ (purple dot-dashed curve), $%
\protect\phi=\protect\pi$ (orange solid curve), and $\protect\phi=3\protect\pi/2$ (cyan dot-dashed curve).}
\end{figure}
The phenomenon of slow and fast light emerges in the system as a result of abnormal dispersion and has recently been theoretically observed in cavity magnomechanics by \cite{slow,slow2,CKB}. 
Investigating a feasible physical configuration capable of switching from fast to slow light and vice versa is of great interest. In the following, we provide a cross-cavity magnomechanical system to examine group delay propagation and slow-to-fast light switching in a single configuration using the relative phase of the two probing fields present in the system.
The group delay of the output field is associated with the phase of the transmitted probe field, which is described as
\begin{eqnarray}
\tau_g&=&\frac{d\Psi}{d \sigma}=\Im[\frac{1}{T_p}\frac{\delta T_p}{\delta\sigma}].
\end{eqnarray}
where $\Psi=\text{arg} [T_p]$. The sign of $\tau_g$ can determine the temporal delay profile. There are two ways to evaluate the group delay. (i) A positive group delay ($\tau_g>0$) is responsible for slow light (subluminal) and (ii) negative group delay ($\tau_g<0$) represents the fast light (superluminal) in the system.
In Fig. 9, we plot the transmission group delay against the normalized detuning for different phase values. It is possible to obtain both positive and negative group delays for different phase values at different detuning values as shown in Fig. 9. In the case when the two probes are in phase, i.e., $\phi=0$, the group delay reaches a positive value of $\tau_g=38\mu$s around the probe detuning $\sigma=\pm0.55\omega_{b}$. However, it does not move down to the fast-light region. In contrast, when the two probes are completely out of phase, i.e., $\phi=\pi$, the group delay reaches a maximum of $\tau_g=20\mu$s at the opacity point and moves down to $\tau_g=-13\mu$s around the detuning  $\sigma=\pm0.6\omega_{b}$.
In the case of $\phi=\pi/2$, the group delay curve move from subluminal ($\tau_g=20\mu$s around $\sigma=-0.55\omega_{b}$.) to superluminal ($\tau_g=-13\mu$s around $\sigma=0.4\omega_{b}$). Interestingly enough, the group delay in the case of $\phi=3\pi/2$ is mirror image symmetric to the case of $\phi=\pi/2$.
It is important to mention here that we obtained a group delay ranging from $\tau_g=45\mu$s to $\tau_g=-13\mu$s employing the experimentally feasible parameters. However, previously pulse variation range $\tau_g=4\mu$s to $\tau_g=-6\mu$s has been reported in \cite{{phase}}, with a different configuration model. Hence, compared to the previous report, here we attain a significantly higher range of group delay.
Therefore, based on phase variation, we provide a more tunable and switchable scheme for controlling a slow-to-fast light transition (and vice versa).

\section{Cconclusions}
In conclusion, we investigate the enhanced absorption and the transmission spectrum in a cross-cavity magnoomechanical system, where both microwave cavities are driven by weak probe fields with different phase angles, and a strong pump field drives the magnon mode. We mainly concentrate on the relative intensity amplitudes and the phase of the applied probe fields on the absorption and transmission spectrum. It is important to note that, in contrast to a single transparency window when the magnon–phonon interaction is absent, the presence of either magnon–phonon or magnon-photon (with cavity-2) interactions results in dual transparency windows.
We then demonstrate how the absorption and transmission spectra can be modulated asymmetrically into absorption and amplification by tuning the relative intensity and the phase of the two probe fields. Furthermore, the transmission of the current system can also be controlled
(either suppressed or enhanced) by tuning the relative phase and amplitude.
The combined effect of the magnon-photon and magnon-phonon couplings, with relative phase, helps to switch from the subluminal to superluminal behavior of the probe field in the current system.
Furthermore, long-lived slow light (with a group delay on the order of milliseconds) can be attained by increasing the relative amplitudes of probe fields. The findings of the current study show that modulating the relative phase of applied probe fields in a cross-cavity magnomechanical system may have significant implications in the storage and transfer of quantum information, and even find useful and potential applications in microwave signal processing and quantum sensing.

\end{document}